\newcommand{\CLASSINPUTtoptextmargin}{0.75in}
\newcommand{\CLASSINPUTbottomtextmargin}{1.1in}
\documentclass[conference]{IEEEtran}
\IEEEoverridecommandlockouts

\usepackage{url}
\usepackage{cite}
\usepackage{subcaption}
\usepackage{amsmath,amssymb,amsfonts}
\usepackage{algorithmic}
\usepackage{graphicx}
\usepackage{textcomp}
\usepackage{xcolor}
\def\BibTeX{{\rm B\kern-.05em{\sc i\kern-.025em b}\kern-.08em
    T\kern-.1667em\lower.7ex\hbox{E}\kern-.125emX}}
\begin{document}

\title{Design of an Edge-based Portable EHR System for Anemia Screening in Remote Health Applications
\thanks{NSF-EPSCoR Center for the Advancement of Wearable Technologies, NSF grant OIA-1849243}
}

\author{
Sebastián A. Cruz Romero$^{1}$,
Misael J. Mercado Hernández$^{2}$,
Samir Y. Ali Rivera$^{3}$,\\
Jorge A. Santiago Fernández$^{4}$,
Wilfredo E. Lugo Beauchamp$^{5}$ \\
\textit{Computer Science and Engineering},\\
\textit{University of Puerto Rico at Mayagüez}, \\ Mayagüez, Puerto Rico \\
$^{1}$sebastian.cruz6@upr.edu,
$^{2}$misael.mercado1@upr.edu,
$^{3}$samir.ali@upr.edu,\\
$^{4}$jorge.santiago32@upr.edu,
$^{5}$wilfredo.lugo1@upr.edu
}

\maketitle

\begin{abstract}
    The design of medical systems for remote, resource-limited environments faces persistent challenges due to poor interoperability, lack of offline support, and dependency on costly infrastructure. Many existing digital health solutions neglect these constraints, limiting their effectiveness for frontline health workers in underserved regions.

    This paper presents a portable, edge-enabled Electronic Health Record platform optimized for offline-first operation, secure patient data management, and modular diagnostic integration. Running on small-form factor embedded devices, it provides AES-256 encrypted local storage with optional cloud synchronization for interoperability.
    
    As a use case, we integrated a non-invasive anemia screening module leveraging fingernail pallor analysis. Trained on 250 patient cases (27\% anemia prevalence) with KDE-balanced data, the Random Forest model achieved a test RMSE of 1.969 g/dL and MAE of 1.490 g/dL. A severity-based model reached 79.2\% sensitivity. To optimize performance, a YOLOv8n-based nail bed detector was quantized to INT8, reducing inference latency from 46.96 ms to 21.50 ms while maintaining mAP@0.5 at 0.995.
        
    The system emphasizes low-cost deployment, modularity, and data privacy compliance (HIPAA/GDPR), addressing critical barriers to digital health adoption in disconnected settings. Our work demonstrates a scalable approach to enhance portable health information systems and support frontline healthcare in underserved regions.
\end{abstract}


\section{Introduction}
Expanding access to essential healthcare services in remote and underserved regions requires more than digitizing records or deploying mobile apps. In areas where stable power supplies, broadband connectivity, and centralized infrastructure are unreliable or entirely absent, conventional Electronic Health Record (EHR) systems and cloud-reliant mobile health (mHealth) solutions fall short. \cite{openehr2024, healthcareaccess2022} The pressing need is for robust, self-sufficient, portable healthcare technologies; systems designed to function effectively under constraints, delivering medical support directly at the point of need, regardless of infrastructure availability. \cite{iotfoghealth2023, mhealthreview2024, rpmintegration2022}

Rather than treating these constraints as limitations, our approach embraces them as core design drivers. We propose a methodology for developing small-scale, edge-enabled health systems that prioritize autonomy, resilience, and adaptability \cite{abdellatif2020edge}. Such systems must operate independently of persistent internet connections, handle on-device data storage and computation, and be deployable on low-power, cost-effective hardware \cite{abdellatif2020ihealth}. They should not only collect health data but actively support screening, triage, and diagnostic decision-making in the field. The shift from centralized data management to decentralized medical support redefines how portable health technologies can empower frontline health workers and underserved communities \cite{nguyen2021bedgehealth}.

In this context, Artificial Intelligence (AI) holds great promise for augmenting clinical capabilities. However, the typical dependency on cloud-computing platforms, high-bandwidth networks, and continuous power presents a mismatch with the realities of resource-constrained environments \cite{phn2025ai, turn0search3}. Current health technologies often prioritize sophisticated architectures ill-suited for off-grid operation, or they reduce mHealth systems to mere data capture interfaces, lacking meaningful integration to clinical triage events or failover robustness \cite{mhealthreview2023}. Addressing this disconnect requires rethinking how AI and healthcare technologies are integrated, not as cloud-tethered services, but as embedded, edge-resident solutions designed for harsh deployment conditions \cite{medaide2024, nguyen2021bedgehealth}.

Certain health conditions, notably anemia, exemplify the global impact of such gaps in medical support infrastructure. Affecting approximately 1.9 billion people worldwide, anemia imposes a significant burden on public health, particularly in low- and middle-income countries \cite{Manish2025-le, Balarajan2011, Kassebaum2014}. Effective anemia management depends on regular screening and monitoring, services often inaccessible in remote or underserved regions. \cite{Alem2023-hl, Ge2025-jx} Thus, anemia serves as a compelling demonstrator use case for validating robust, portable health support systems. A solution capable of supporting anemia screening in disconnected environments inherently addresses broader challenges of rural diagnostics and decentralized health services.

The adoption of Edge AI and embedded computing platforms offers a pathway toward realizing this vision, justified by the importance of making reliable sensing technologies to acquire large amounts of representative streams of data. \cite{gill2024edgeaitaxonomysystematic, SINGH202371} By localizing computation and data handling, edge-based systems minimize reliance on external infrastructure, lower operational costs, and enhance data privacy. Coupled with the imperative to reduce medical device e-waste and develop sustainable technologies, edge computing aligns with global health and environmental goals \cite{Ongaro2022-au, who2023, bressler2023}. The availability of low-power devices and open-source software frameworks enable the creation of scalable, equitable health technologies tailored to the needs of underserved populations.

In this work, we present a portable, edge-enabled health support platform designed from the ground up to operate in disconnected, resource-limited settings. Our system integrates a non-invasive anemia screening module as a representative case, but the underlying architecture emphasizes generalizable robustness and modularity. Specifically, we developed:

\begin{enumerate}
    \item A joint-model framework for fingernail detection and hemoglobin estimation,
    \item Secure and privacy-focused EHR system hosted locally on battery-powered microprocessing unit (MPU).
    \item Mobile-friendly web application for managing, interpreting, and visualizing EHRs and screening results.
\end{enumerate}

\section{Methodology}

\subsection{System Architecture and Design Rationale}

We developed a portable, edge-enabled health platform designed for resource-constrained environments where reliable power, connectivity, and IT infrastructure are limited. The system integrates:

\begin{enumerate}
    \item a non-invasive anemia screening module,
    \item a local EHR manager, and
    \item a frontend interface for healthcare workers.
\end{enumerate}

Unlike cloud-dependent solutions, the platform follows an offline-first architecture. All data processing, storage, and diagnostic inference occur locally on-device, ensuring autonomous operation. Cloud synchronization is supported but strictly optional. This design choice ensures reliability in disconnected and low-infrastructure settings.

RESTful APIs enable modular communication between components (e.g., medical screening models, local database, and frontend) ensuring decoupling, maintainability, and scalability. The system is deployed on low-power embedded hardware ($\leq$10W power consumption and thermal output), such as the NVIDIA Jetson Nano \cite{nvidia_jetson_orin_nano_ds}, balancing computational needs with energy efficiency.

Key design drivers included:

\begin{itemize}
    \item power scarcity necessitating edge computing on efficient hardware.
    \item unreliable connectivity mandating local data handling, and
    \item lack of specialized IT support requiring user-friendly, robust systems.
\end{itemize}

To meet these, we favored lightweight ML models (Random Forest for hemoglobin estimation) over deep networks, due to their lower compute overhead and explainability. Model quantization and other optimizations were applied to a larger deep learning architecture to reduce latency and power usage, enabling real-time inference at the edge.

For data privacy compliance, we employed record-level AES-256 \cite{NISTFIPS197upd1} encryption within the local PostgreSQL database, ensuring secure data at rest and Role-Based Access Control (RBAC) governs access. The system’s frontend interface, built with ReactJS and Fast APIs, supports intuitive data entry, result visualization, and trend tracking. Offline caching ensures uninterrupted functionality.

\begin{figure*}[ht]
    \centering
    \includegraphics[width=\linewidth]{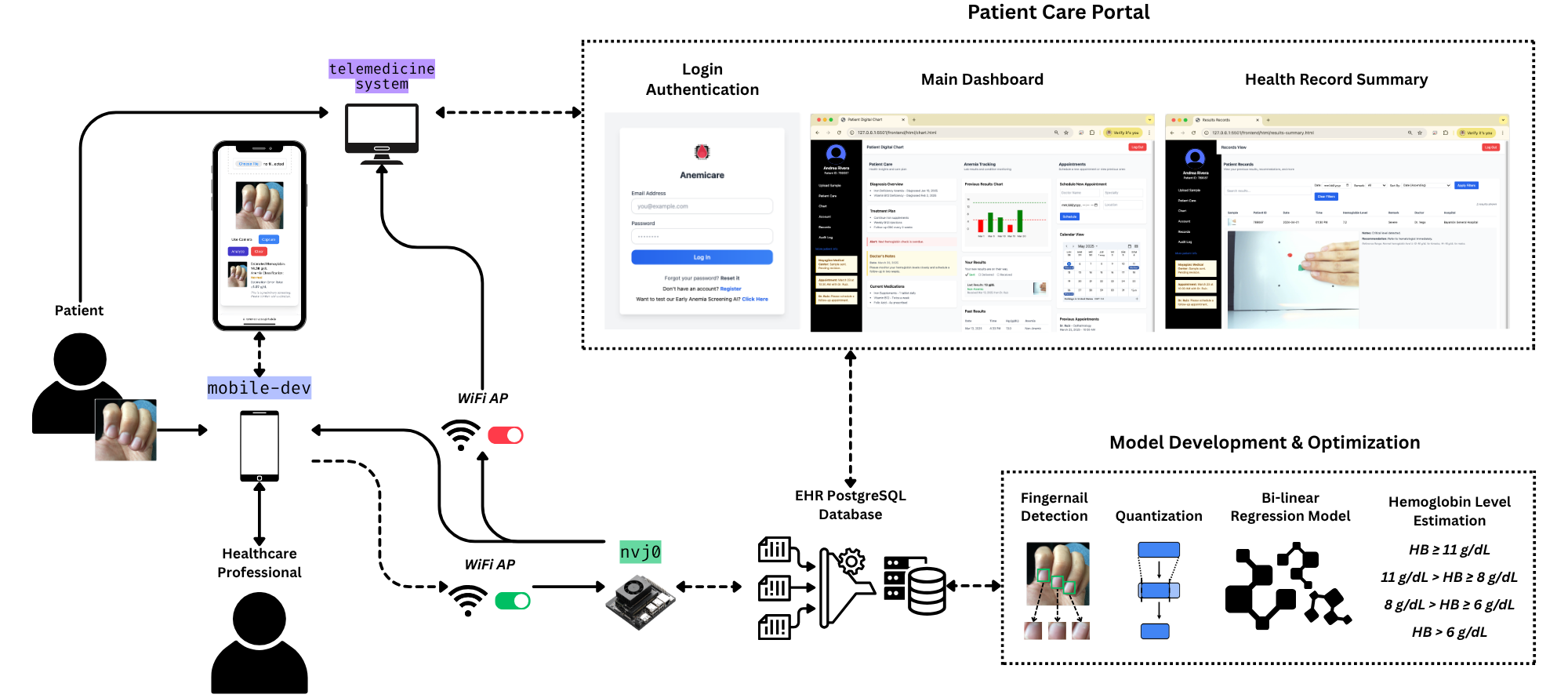}
    \caption{System architecture for our patient monitoring and clinical records platform, integrating edge and smart devices with a front-end dashboard.}
    \label{fig:system_overview}
\end{figure*}

This architecture, illustrated in Figure~\ref{fig:system_overview} reflects a constraint-driven, robustness-first approach to mHealth system design, enabling scalable deployment of diagnostic and health record solutions where they are needed most.

\subsection{Anemia Screening Module}
\subsubsection{Dataset and Preprocessing}
\begin{figure}
    \centering
    \includegraphics[width=\linewidth]{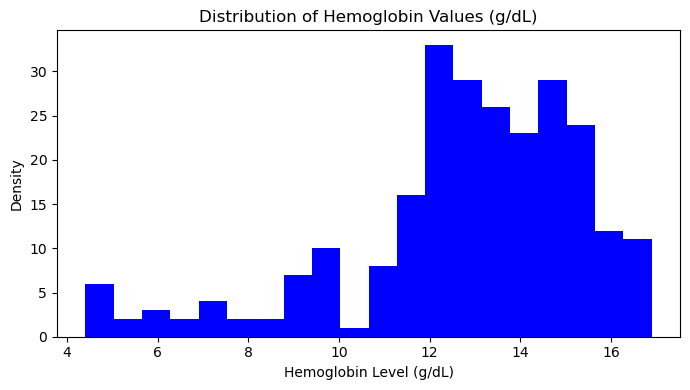}
    \caption{Histogram showing the empirical distribution of hemoglobin levels (g/dL) as measured using an FDA-certified clinical laboratory method. The dataset demonstrates a right-skewed distribution, with the majority of observations clustering between 11 and 15 g/dL, reflecting the natural prevalence of non-anemic subjects in the sampled population.}
    \label{fig:hblevel_dist}
\end{figure}

The anemia screening module is built upon a publicly available dataset comprising 250 fingernail images, each annotated with clinical hemoglobin levels obtained through laboratory testing. \cite{yakimov2024dataset} The dataset reflects a realistic class imbalance, showed in Figure~\ref{fig:hblevel_dist}, with approximately 27\% of subjects classified as anemic (Hb $<$ 12 g/dL), posing challenges for model training and evaluation.

Initial preprocessing involved Region of Interest (ROI) extraction, isolating the fingernail area from each image. This was achieved using a YOLOv8n object detection model, trained on a subset of manually annotated images to detect nailbed regions. Images were subjected to color normalization to mitigate variations due to lighting and skin tone. Statistical features were computed from the RGB and L*a*b* color spaces, including measures of central tendency, dispersion, and skewness. These features capture the visual pallor associated with anemia, which is a clinically recognized diagnostic indicator. The preprocessing pipeline was designed to be fully automated, ensuring reproducibility and scalability for field deployment.

\begin{figure}[ht]
    \centering
    \begin{subfigure}[b]{1\linewidth}
        \centering
        \includegraphics[width=\linewidth]{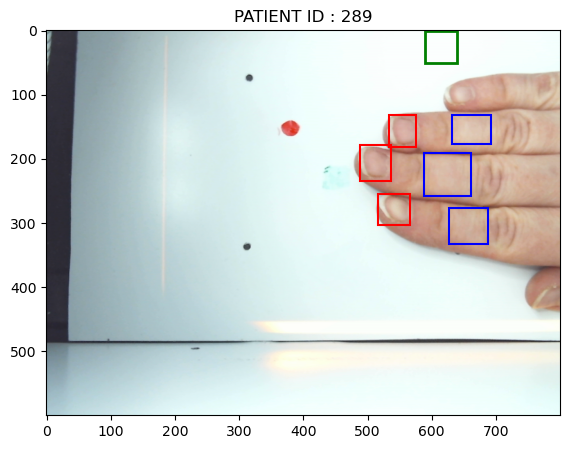}
        \caption{Image samples with marked bounding boxes in regions of interests. Red boxes denote nail beds and blue boxes skin tissue. Green boxes where used as a reference for image normalization.}
        \label{fig:kdebal_remark}
    \end{subfigure}
    \hfill
    \begin{subfigure}[b]{\linewidth}
        \centering
        \includegraphics[width=\linewidth]{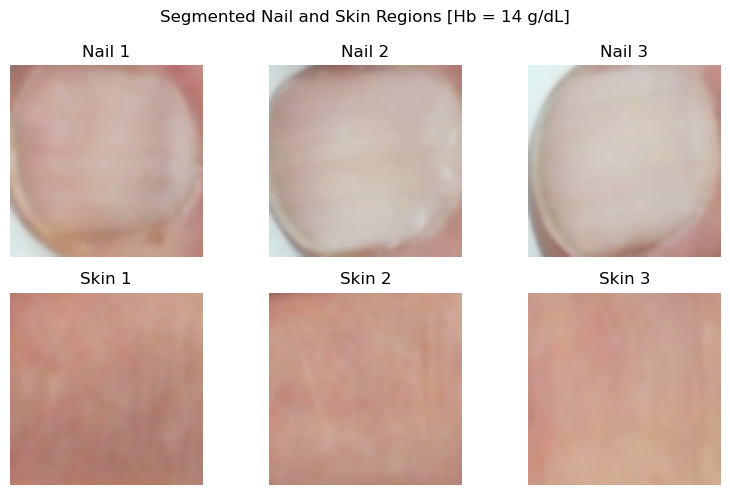}
        \caption{Cropped image sample of nail beds and skin for a patient with a hemoglobin level of 14 g/dL.}
        \label{fig:kdebal_severity}
    \end{subfigure}
    \caption{Image samples from Yakimov et. al. \cite{yakimov2024dataset} including a original image samples, nail and skin bounding boxes with a white reference region, and cropped nail and skin regions.}
\end{figure}

\subsection{Model Development \& Optimization}
For hemoglobin estimation, multiple regression models were evaluated, including Random Forest, Gradient Boosting, Support Vector Regression (SVR), ElasticNet, Ridge, Lasso, RANSAC, and Huber regressors. Model selection criteria emphasized not only predictive accuracy but also computational efficiency and interpretability. Given the dataset's limited size and the deployment constraints, a Random Forest Regressor was selected for its robustness to noise, ability to model non-linear relationships, and low inference complexity.

\begin{figure}[ht]
    \centering
    \begin{subfigure}[b]{\linewidth}
        \centering
        \includegraphics[width=\linewidth]{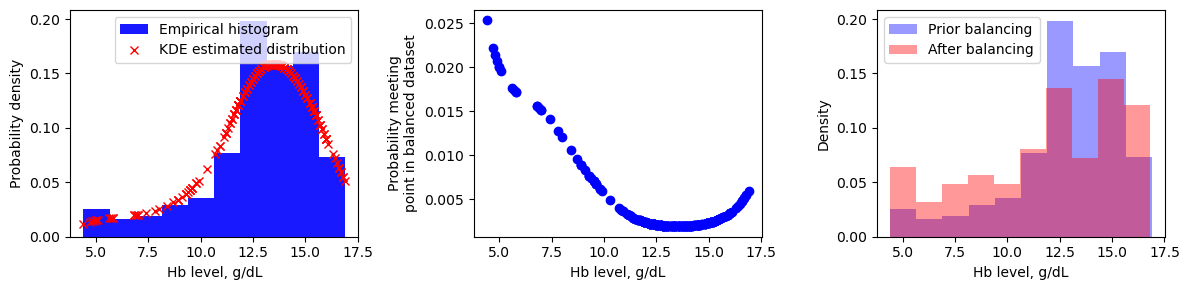}
        \caption{KDE-balanced data by remark class}
        \label{fig:kdebal_remark}
    \end{subfigure}
    \hfill
    \begin{subfigure}[b]{\linewidth}
        \centering
        \includegraphics[width=\linewidth]{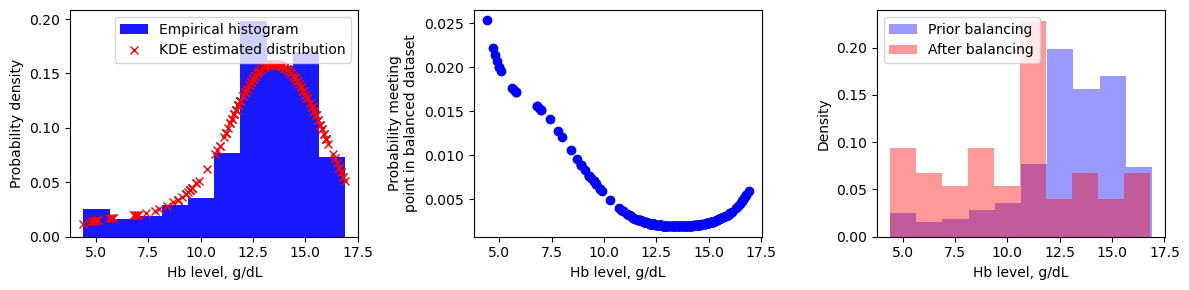}
        \caption{KDE-balanced data by severity class}
        \label{fig:kdebal_severity}
    \end{subfigure}
    \caption{Kernel Density Estimation (KDE) was used to rebalance the training data by redistributing samples across "remark" (anemic vs. non-anemic) and "severity" (non-anemic, mild, moderate, severe) labels. This reweighting mitigated class imbalance and strengthened the learning signal for underrepresented categories. Overlayed histograms (blue: original, pink: KDE-reweighted) illustrate the suppression of overrepresented ranges and amplification of underrepresented ones across the hemoglobin spectrum.}
\end{figure}

To address the dataset's inherent class imbalance, Kernel Density Estimation (KDE) based undersampling was applied, shown in Figure~\ref{fig:kdebal_remark} and~\ref{fig:kdebal_severity}. This approach improved model generalization and prevented bias towards the majority class. The Random Forest model was trained with 100 trees, a maximum depth of 10, and mean squared error as the loss function. Cross-validation (7-fold) was employed to mitigate overfitting and ensure robust performance evaluation. Moreover, model compression techniques were applied to optimize for edge deployment.

\subsection{EHR Management Module}

The EHR management module was designed to provide secure, local-first health data management, with optional cloud synchronization for extended interoperability. At its core, the module utilizes a PostgreSQL database for on-device storage, chosen for reliability and suitability for telemedicine systems. To ensure data security, all patient records are encrypted using the Advanced Encryption Standard with a 256-bit key (AES-256) \cite{NISTFIPS197upd1}, aligning with industry standards for health information protection under the Health Insurance Portability and Accountability Act (HIPAA) \cite{hipaa2023} and the General Data Protection Regulation (GDPR) \cite{gdpr2016}.

Access to EHR data is governed by a Role-Based Access Control (RBAC) system, which authenticates users and enforces data access permissions. This ensures that only authorized personnel can view, modify, or synchronize patient records. The encryption and access control mechanisms were designed to operate with minimal performance overhead, with benchmark tests indicating encryption/decryption latency of less than 30 ms per record on the Jetson Nano platform.


\subsection{Frontend Interface}
A web-based dashboard interface was developed to facilitate interaction between healthcare workers and the system. Built using ReactJS for the frontend and a lightweight FastAPI backend API, the interface allows users to:

\begin{itemize}
    \item Register and manage patient records.
    \item Visualize anemia screening results in real-time.
    \item Track longitudinal health trends for individual patients.
    \item Export anonymized data for reporting and analytics.
\end{itemize}

The interface is designed with a responsive layout, ensuring usability across devices with varying screen sizes, including tablets and smartphones commonly used in field clinics. Offline operation is supported through local caching mechanisms, enabling data access even in the absence of network connectivity.

\begin{figure}
    \centering
    \includegraphics[width=.90\linewidth]{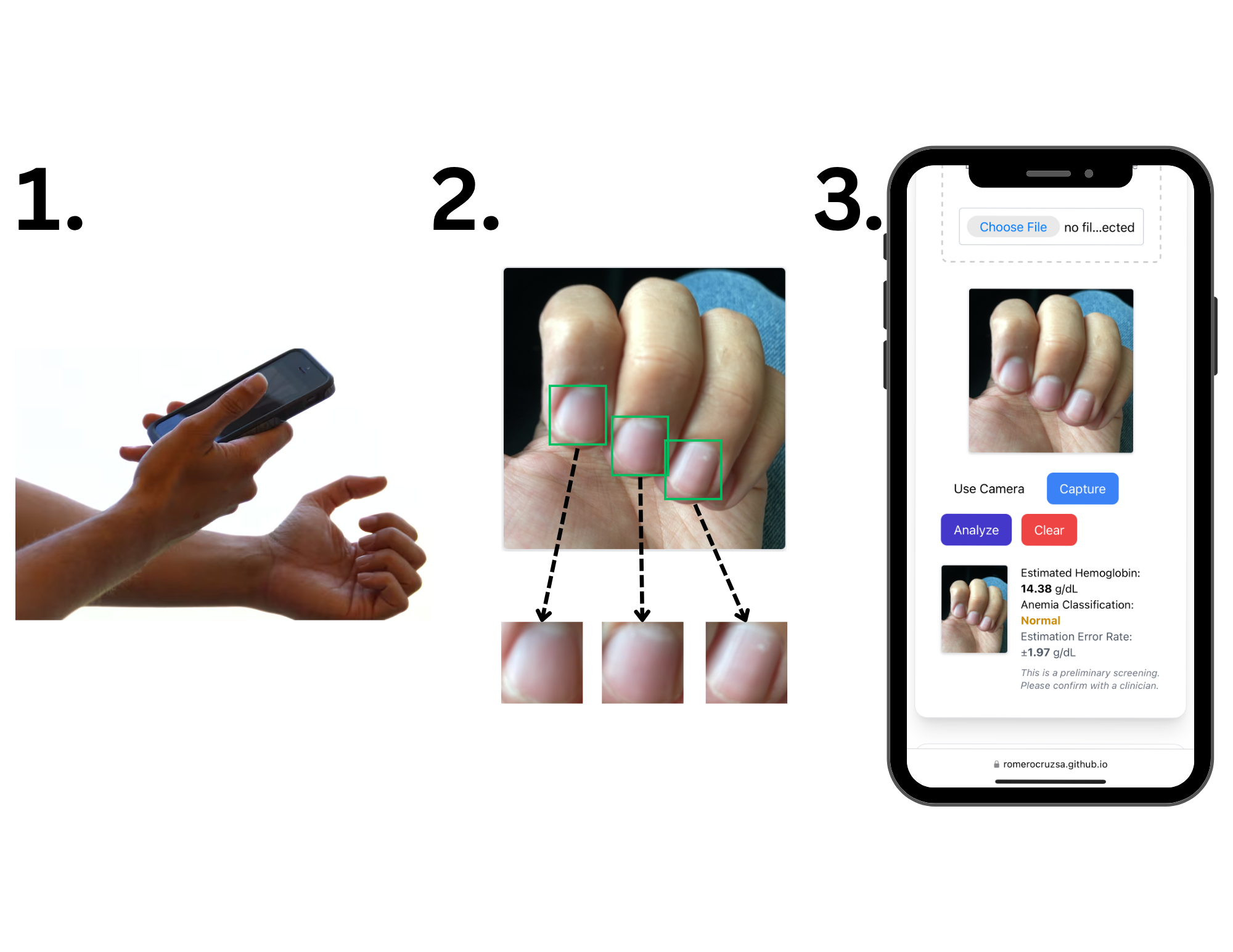}
    \caption{Step-by-step protocol for anemia screening}
    \label{fig:ui-ux_view}
\end{figure}

\begin{figure}
    \centering
    \includegraphics[width=0.90\linewidth]{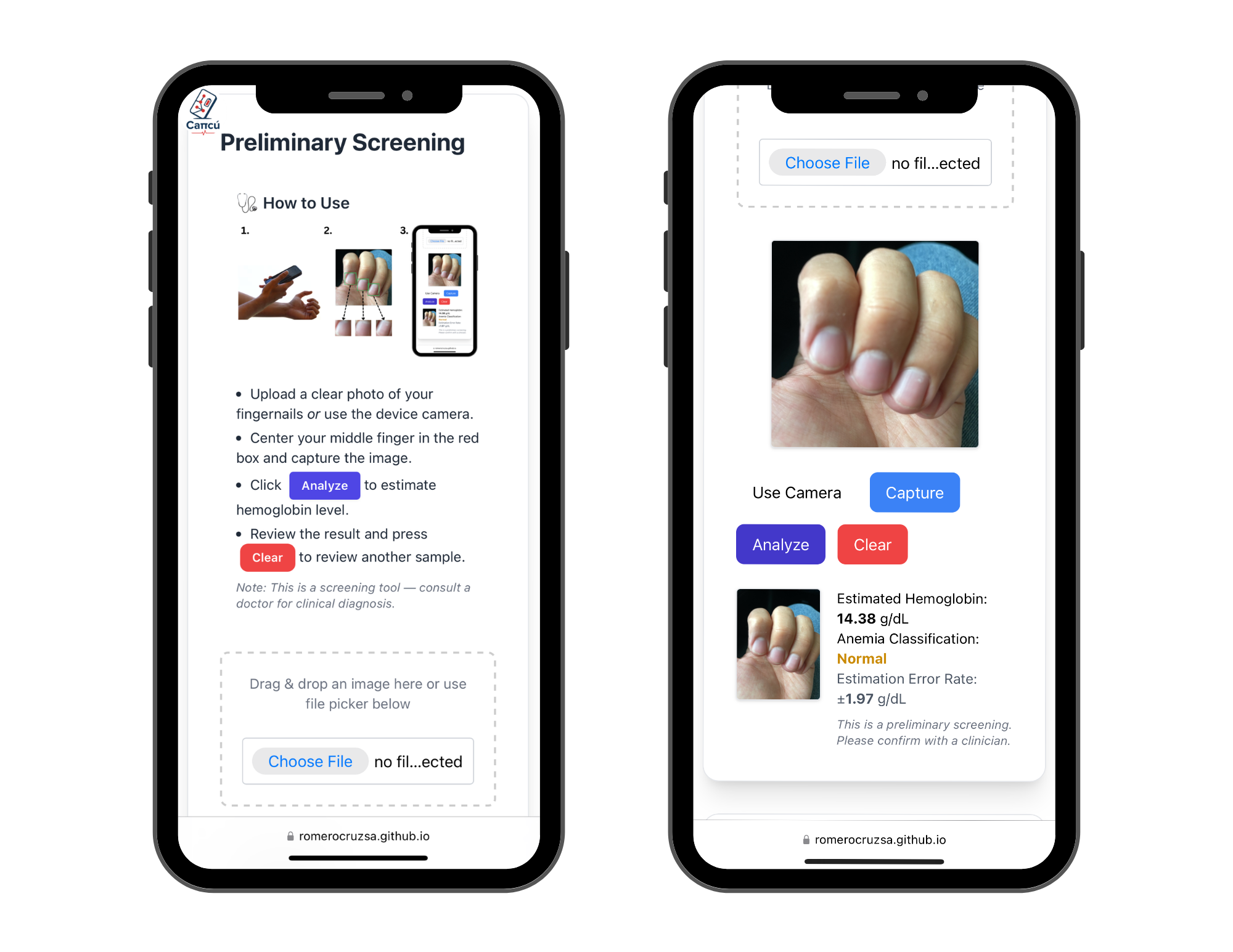}
    \caption{Single-use app view for quick intervention}
    \label{fig:single-use_view}
\end{figure}

\begin{figure*}[htbp]
    \centering
    \begin{subfigure}[b]{0.45\textwidth}
        \includegraphics[width=\linewidth]{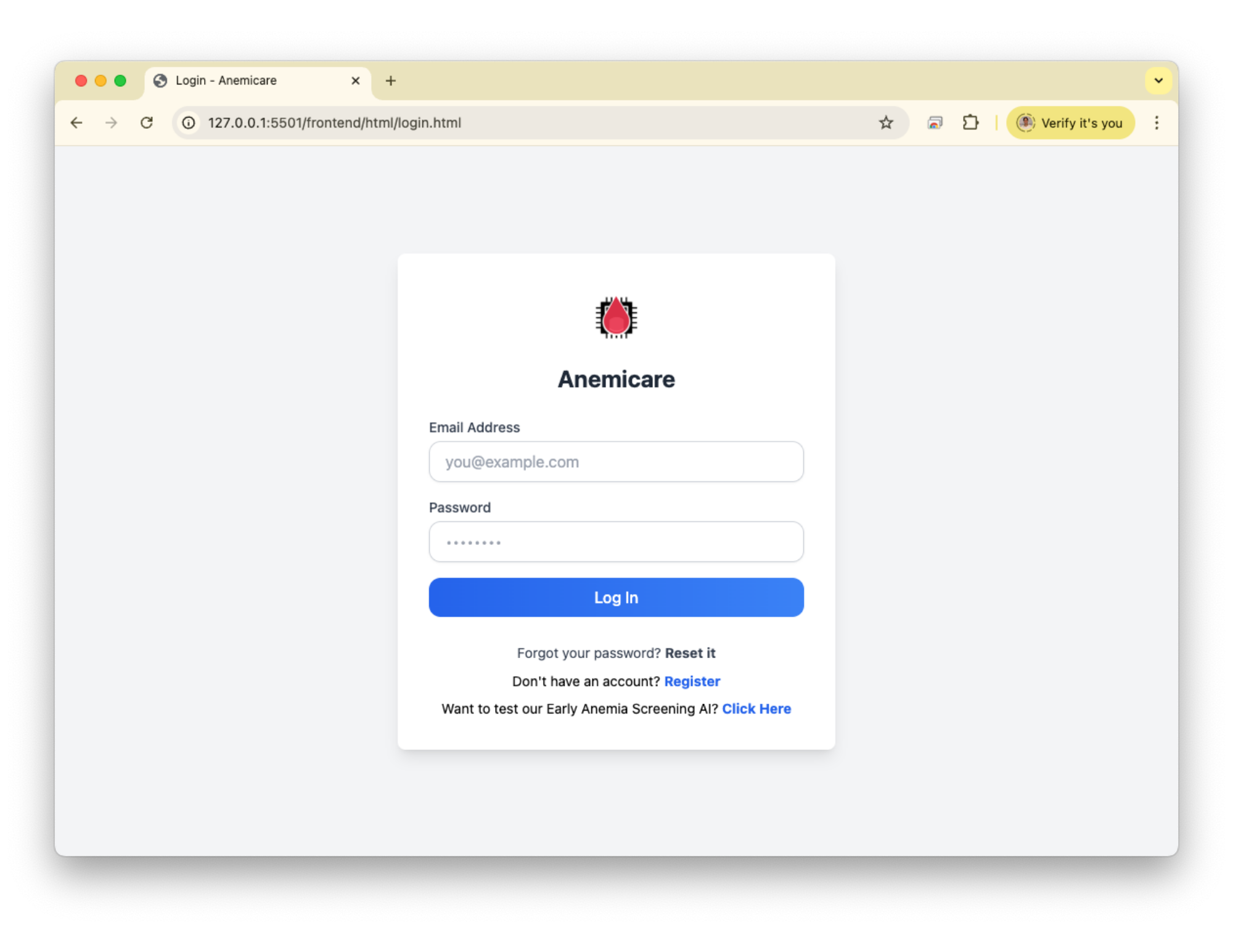}
        \caption{Login screen for user authentication in the Point-of-Care system.}
        \label{fig:login_page}
    \end{subfigure}
    \hfill
    \begin{subfigure}[b]{0.45\textwidth}
        \includegraphics[width=\linewidth]{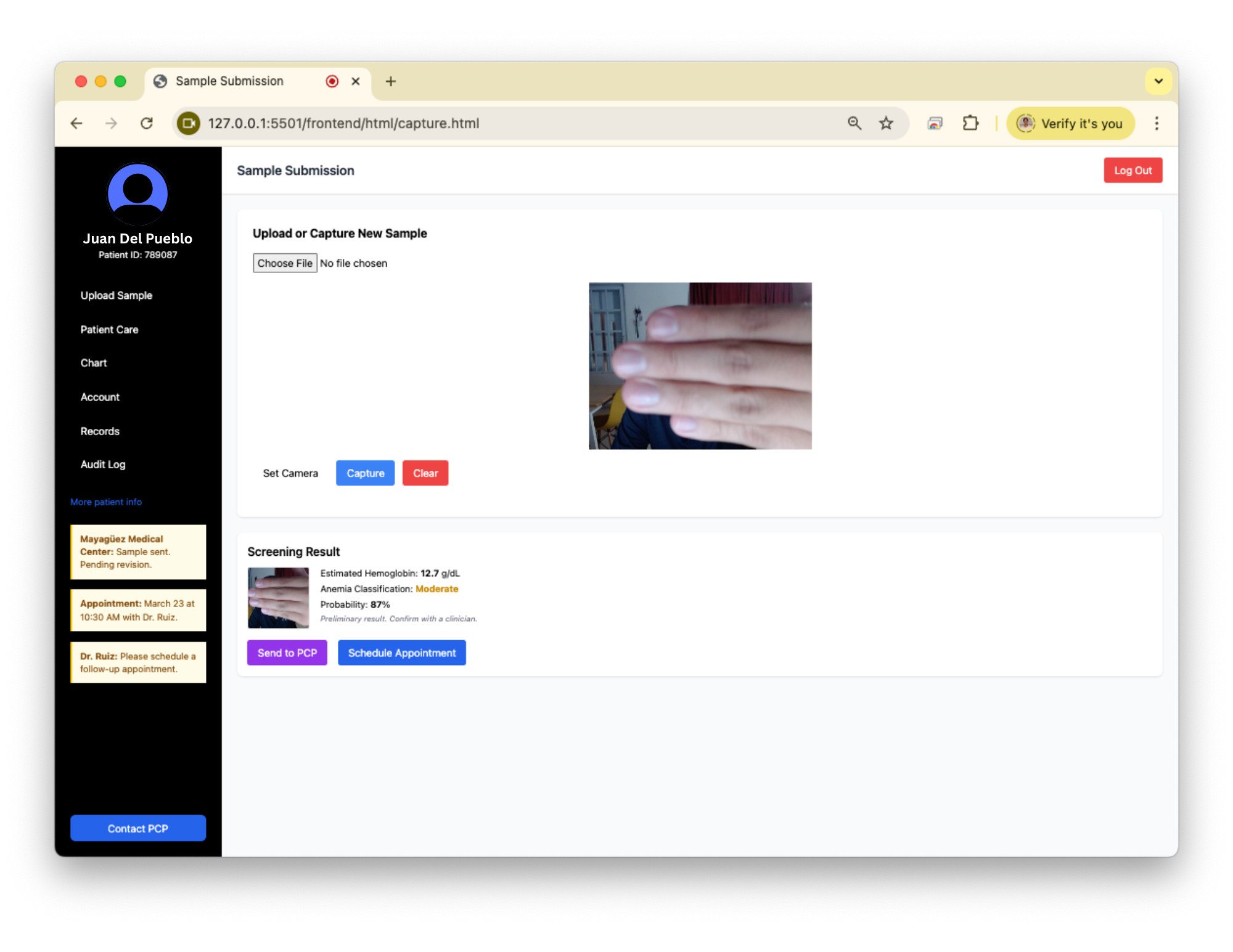}
        \caption{Digital patient chart view showing results, treatment plan, appointments, and historical data.}
        \label{fig:sample_submission}
    \end{subfigure}

    \vspace{1em} 

    \begin{subfigure}[b]{0.45\textwidth}
        \includegraphics[width=\linewidth]{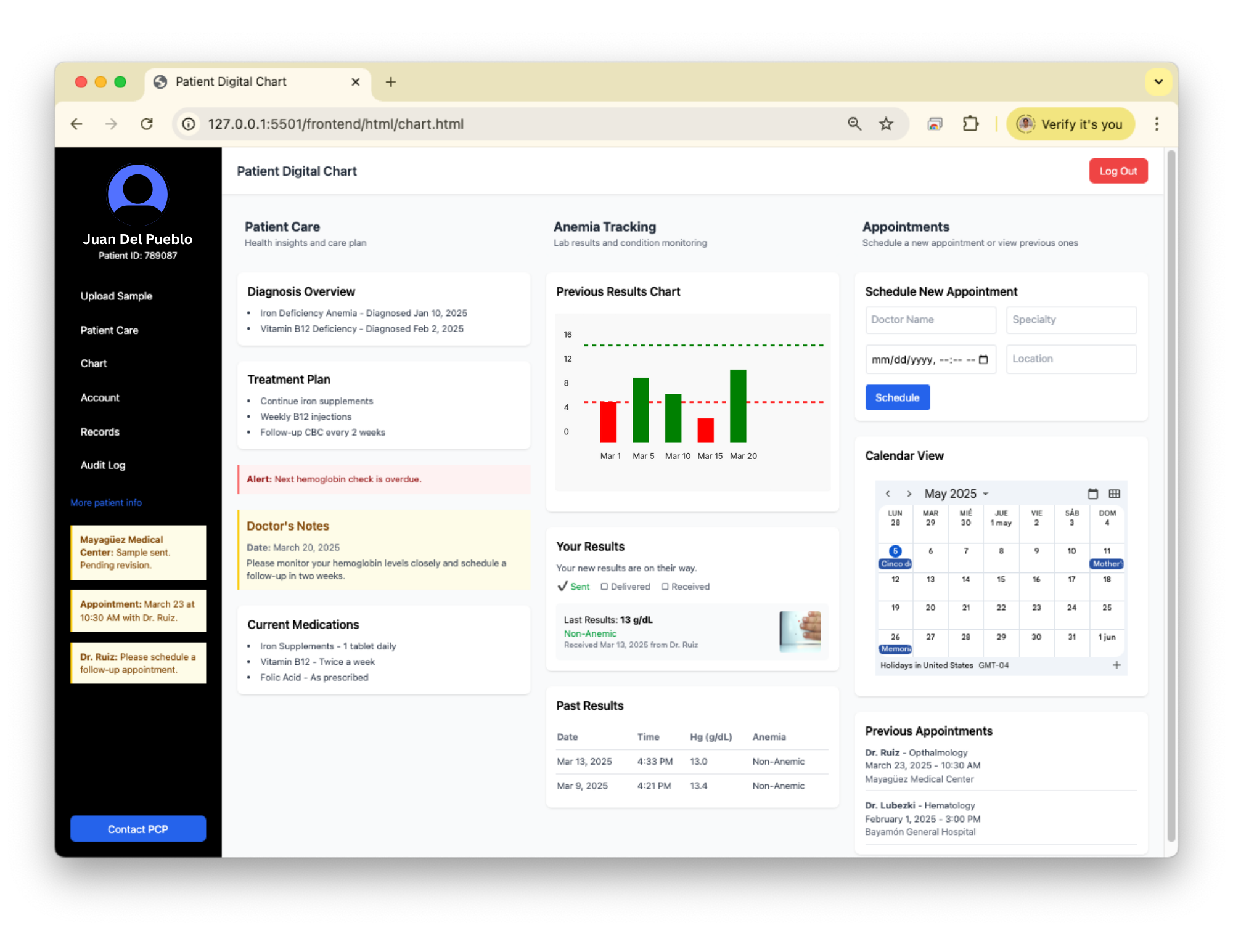}
        \caption{Patient records view with hemoglobin test results and visual analysis.}
        \label{fig:records_summary}
    \end{subfigure}
    \hfill
    \begin{subfigure}[b]{0.45\linewidth}
        \includegraphics[width=\linewidth]{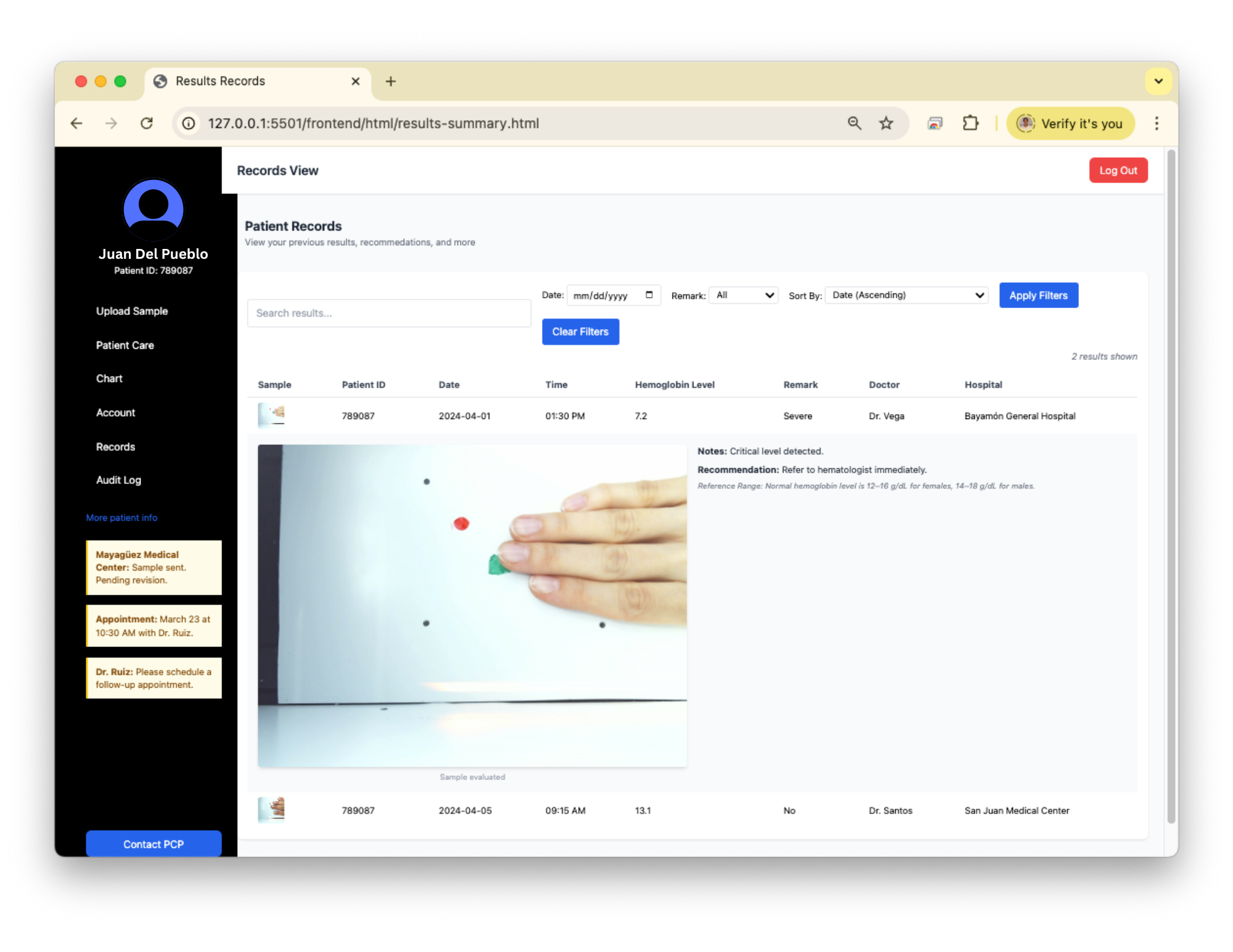}
        \caption{Sample submission interface with real-time camera capture and AI-generated screening result.}
        \label{fig:records_view}
    \end{subfigure}

    \caption{Key interfaces of the Point-of-Care system, including login, patient records, chart overview, and sample submission with AI-based anemia screening.}
    \label{fig:grid}
\end{figure*}

\subsection{Hardware Deployment \& Benchmarking}
The fully integrated system was deployed on NVIDIA Jetson Orin Nano \cite{nvidia_jetson_orin_nano_ds} platforms for performance evaluation. Inference latency and power consumption were measured under typical usage scenarios, including continuous screening and batch data synchronization. A P3 P4400 Kill-A-Watt Electricity Usage Monitor \cite{p3_killawatt_p4400} was used to measure power consumption by connecting the Jetson Nano through the monitor.

On the Jetson Nano, the anemia screening pipeline achieved an average inference latency of 28 ms per image, with a total average end-to-end response time (including data storage and UI update) of approximately 42 to 58 ms per patient. Power consumption remained under 7W during active inference cycles. These results validate the system's ability to deliver real-time diagnostic support and health data management in field conditions, aligning with the project's design objectives and constraints.

\section{Results \& Discussion}

To evaluate the proposed portable EHR platform with integrated anemia screening, we conducted a series of model performance benchmarks, edge device deployment tests, and end-to-end system validation. The evaluation focused on measuring diagnostic accuracy, computational efficiency, deployment feasibility, and operational robustness in resource-constrained settings.

\subsection{Anemia Detection Model Performance}
\subsubsection{\textbf{Fingernail Detection}}
We used the YOLOv8n object detection model, trained on manually annotated nailbed bounding boxes such that we can accurately identify nail images by their location from the hand and convex-end shape of the nail bed. To ensure feasibility for edge deployment, we performed Post-Training Quantization (PTQ) \cite{gholami2021}, reducing the bit-width for computing weights and biases to integer arithmetic. We reduced inference latency by 54.2\% (from 46.96 ms to 21.50 ms) while preserving detection accuracy (mAP@0.5 remained at 0.995). This effectively reduced model size by approximately 50\% of the floating point (FP32) model baseline.

\begin{table}[ht]
    \centering
    \caption{Comparison of YOLOv8n model performance before and after INT8 Post-Training Quantization.}
    \label{tab:yolo_quant_comparison_transposed}
    \begin{tabular}{|l|c|c|}
    \hline
    \textbf{Metric} & \textbf{FP32 Baseline} & \textbf{INT8 Quantized} \\
    \hline
    Precision & 0.993 & 0.999 \\
    Recall & 1.000 & 1.000 \\
    mAP@0.5 & 0.995 & 0.995 \\
    mAP@0.5:0.95 & 0.692 & 0.638 \\
    Model Size & 6.2 MB & 6.2 MB \\
    Inference Speed & 46.96 ms & 21.50 ms \\
    \hline
    \end{tabular}
\end{table}

\subsubsection{\textbf{Hemoglobin Level Estimation}}
Eight baseline models were evaluated across both classification and regression tasks: ElasticNet, Ridge, Lasso, Random Forest, Gradient Boosting, Support Vector Regression (SVR), RANSAC Regressor, Huber
Regressor. This diversity was intentional to evaluate how different modeling assumptions, particularly linearity, collinearity, and model interpretability, affect clinical performance. Given the high degree of feature collinearity in medical image-derived descriptors (e.g., overlapping RGB percentiles and texture scores) the Random Forest regressor achieved the best RMSE of 1.969 g/dL when evaluated on KDE-balanced data by remark (e.g., anemic, non-anemic).

Table~\ref{tab:results_remark} summarizes the results (Remark-based evaluation).

\begin{table}[ht]
    \centering
    \caption{Performance summary of regression models on KDE balanced data by remark ordered by lowest test set RMSE.}
    \label{tab:results_remark}
    \begin{tabular}{|l|c|c|c|c|}
    \hline
    \textbf{Model} & \textbf{Sensitivity} & \textbf{Specificity} & \textbf{MAE} & \textbf{RMSE} \\
    \hline
    Random Forest & 0.360 & 0.784 & 1.490 & 1.969 \\
    Gradient Boosting & 0.440 & 0.800 & 1.626 & 2.132 \\
    SVR & 0.560 & 0.720 & 1.652 & 2.147 \\
    ElasticNet & 0.560 & 0.720 & 1.655 & 2.154 \\
    Lasso & 0.560 & 0.720 & 1.658 & 2.158 \\
    RANSAC & 0.560 & 0.744 & 1.657 & 2.161 \\
    Ridge & 0.560 & 0.744 & 1.681 & 2.184 \\
    Huber & 0.520 & 0.736 & 1.686 & 2.193 \\
    \hline
    \end{tabular}
\end{table}

\begin{figure}[ht]
    \centering
    \begin{subfigure}[b]{\linewidth}
        \centering
        \includegraphics[width=\linewidth]{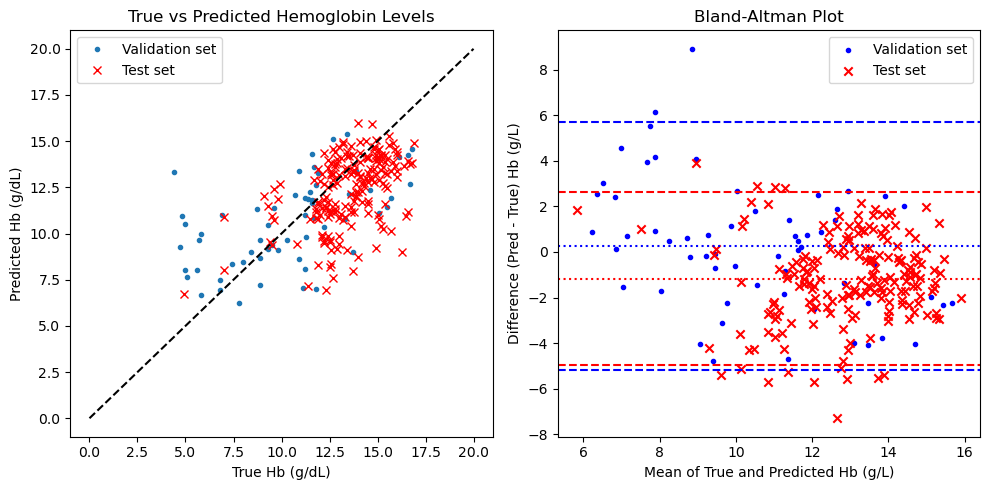}
        \caption{Scatter plot comparing true versus predicted hemoglobin values on the validation (blue) and test (red) sets, with a dotted identity line (y = x) indicating perfect prediction. The accompanying Bland-Altman plot assesses agreement between predicted and true values, showing acceptable bias and limits of agreement across both datasets. These results support the model’s suitability for population-level screening.}
        \label{fig:kdebal_remark}
    \end{subfigure}
    \hfill
    \begin{subfigure}[b]{\linewidth}
        \centering
        \includegraphics[width=\linewidth]{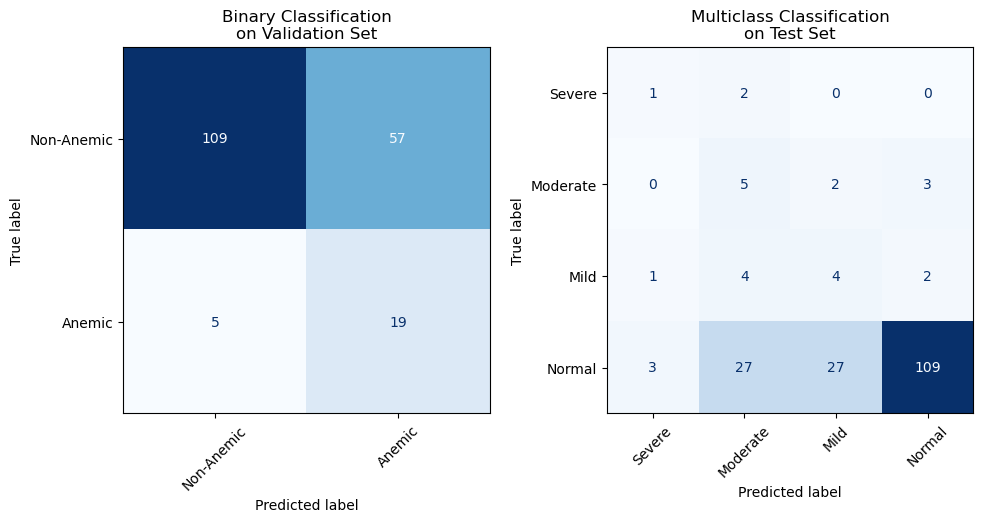}
        \caption{Confusion matrices for binary (validation) and multiclass (test) classification tasks based on predicted anemia status and severity.}
        \label{fig:kdebal_severity}
    \end{subfigure}
    \caption{The best-performing model (Random Forest) captures the overall trend but slightly underestimates higher Hb levels in both validation and test sets. In classification, it performs well for non-anemic cases but struggles with accurate detection of mild and moderate anemia.}
\end{figure}

We also surveyed models to see their performance when data was balanced by severity (e.g., non-anemic, mild, moderate, severe) classes. For clinical triage, severity-based classification was prioritized. Random forest delivered 79.2\% sensitivity with a manageable RMSE of 2.264 g/dL. Table~\ref{tab:results_remark} summarizes the results (Remark-based evaluation). 

\begin{table}[ht]
        \centering
    \caption{Performance summary of regression models on KDE balanced data by severity ordered by lowest test set RMSE.}
    \label{tab:results_severity}
    \begin{tabular}{|l|c|c|c|c|}
    \hline
    \textbf{Model} & \textbf{Sensitivity} & \textbf{Specificity} & \textbf{MAE} & \textbf{RMSE} \\
    \hline
    Random Forest & 0.792 & 0.657 & 1.791 & 2.264 \\
    Gradient Boosting & 0.792 & 0.657 & 1.800 & 2.313 \\
    SVR & 0.833 & 0.596 & 1.999 & 2.473 \\
    ElasticNet & 0.917 & 0.536 & 2.022 & 2.549 \\
    Ridge & 0.917 & 0.536 & 2.026 & 2.572 \\
    Lasso & 0.917 & 0.548 & 2.027 & 2.580 \\
    Huber & 0.917 & 0.506 & 2.092 & 2.669 \\
    RANSAC & 0.875 & 0.494 & 2.189 & 2.764 \\
    \hline
    \end{tabular}
\end{table}

\begin{figure}[ht]
    \centering
    \begin{subfigure}[b]{\linewidth}
        \centering
        \includegraphics[width=\linewidth]{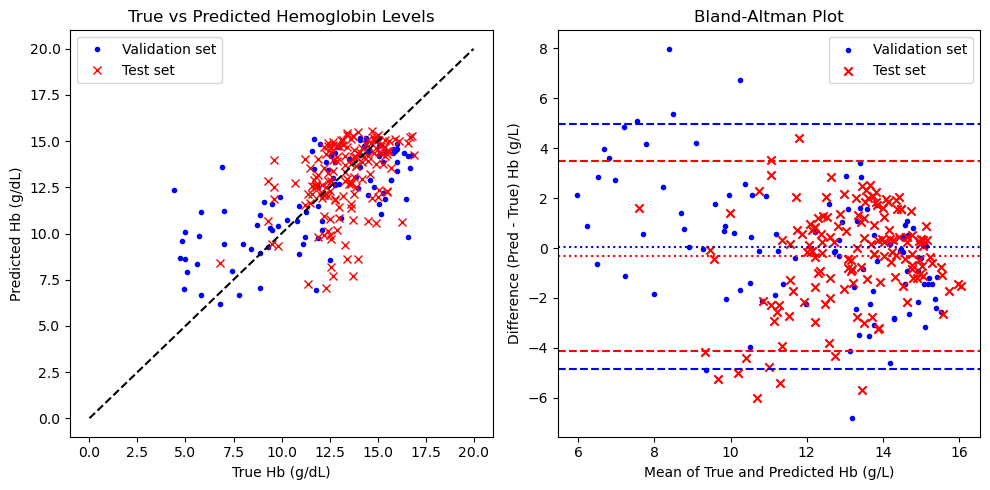}
        \caption{Scatter plot of predicted vs. actual hemoglobin levels on validation (blue) and test (red) sets after KDE-based rebalancing by anemia severity. The Bland-Altman plot shows reduced dispersion and tighter agreement limits, particularly in clinically relevant anemic ranges.}
        \label{fig:kdebal_remark}
    \end{subfigure}
    \hfill
    \begin{subfigure}[b]{\linewidth}
        \centering
        \includegraphics[width=\linewidth]{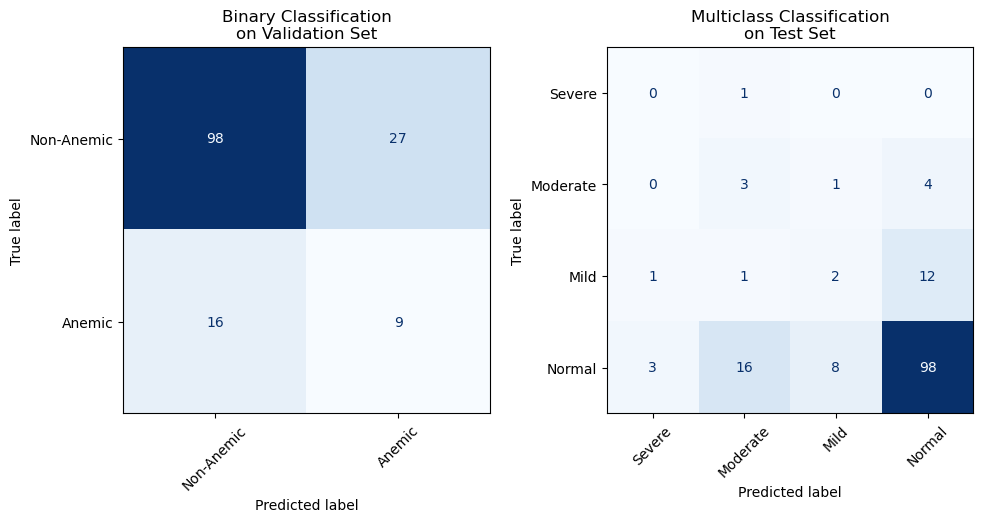}
        \caption{Confusion matrices showing binary classification on validation and multiclass predictions on the test set.}
        \label{fig:kdebal_severity}
    \end{subfigure}
    \caption{The severity-balanced model shows improved linearity compared to the remark-balanced version, especially in the moderate-to-low Hb range. Enhanced representation during training leads to better sensitivity for detecting mild and moderate anemia.}
\end{figure}

\subsection{Data Security \& Privacy Overhead}

Given the sensitivity of patient health data, the proposed EHR platform was implemented to ensure compliance with privacy regulations such as HIPAA in the United States and GDPR in the European Union. We assessed the risks of unauthorized access in case of device loss, theft, or compromise. To mitigate these risks, we implemented Advanced Encryption Standard (AES) with 256-bit keys (AES-256) for securing all locally stored patient records.

\subsubsection{\textbf{Encryption Design \& Integration}}

We opted for a record-level encryption approach within the local PostgreSQL database, ensuring that each patient's health data is individually protected. This granularity allows selective access control and minimizes potential exposure in the event of data breaches. A lightweight cryptographic library optimized for ARM architectures was used to maintain computational efficiency on edge devices.

For authenticated access, a Role-Based Access Control (RBAC) scheme was implemented. Only authorized users (e.g., healthcare workers) with verified credentials could decrypt and access patient records. Furthermore, the system enforced AES-CBC (Cipher Block Chaining) mode with PKCS7 padding, combined with SHA-256 hashing for integrity verification of encrypted data blocks.

\subsection{Discussion}
The results of this work demonstrate the feasibility and effectiveness of a constraint-driven design methodology for developing portable, edge-enabled health support systems tailored to resource-limited environments. By explicitly addressing real-world constraints such as power scarcity, unreliable connectivity, and limited infrastructure, our approach shifts the focus from traditional cloud-dependent health IT solutions to self-sufficient, decentralized systems capable of delivering actionable medical support at the point of need. 

The anemia screening module, while used as a demonstrator in this study, validated the system’s capacity to integrate non-invasive diagnostic capabilities in a compact, computationally efficient manner. Achieving an RMSE of 1.969 g/dL for hemoglobin estimation and 79.2\% sensitivity in severity-based screening confirms that acceptable clinical screening performance can be attained on low-power edge devices. Model optimization techniques, including INT8 quantization and KDE-based balancing, were instrumental in ensuring this balance between accuracy and deployability.

Moreover, the successful implementation of record-level AES-256 encrypted EHR storage and offline-first architecture addresses critical concerns around data privacy, sovereignty, and system resilience.

\section{Conclusion \& Recommendations}

This paper presented the design, development, and evaluation of a portable, edge-enabled EHR platform with integrated anemia screening, engineered specifically for deployment in disconnected, resource-limited healthcare environments. By embracing constraints as design drivers, the proposed system achieves a balance between diagnostic accuracy, computational efficiency, data security, and operational robustness. The demonstrated results highlight the viability of delivering scalable, sustainable, and equitable health technologies that extend medical support directly to where it is most needed, regardless of infrastructure availability.

\subsubsection{\textbf{Broader Impact \& Scalability}}

We contributed to a replicable blueprint for designing robust, small-scale health support systems that can be rapidly deployed in diverse contexts, including:

\begin{itemize}
    \item Rural health clinics lacking stable infrastructure.
    \item Emergency response units in disaster-stricken areas.
    \item Mobile health initiatives by NGOs and public health agencies.
\end{itemize}

By leveraging existing hardware and open-source software stacks, the platform offers a cost-effective, scalable solution for extending healthcare access in underserved regions. The modular architecture further allows integration of additional diagnostic modules (e.g., vital signs monitoring, skin lesion analysis), enhancing the system’s versatility without necessitating fundamental redesign.

From a global health perspective, this approach supports efforts to achieve Universal Health Coverage (UHC) by bridging the technological gap for frontline healthcare workers, enabling them to deliver data-driven, evidence-based care in challenging environments.

\subsubsection{\textbf{Limitations}}

While the results are promising, several limitations warrant acknowledgment:

\begin{enumerate}
    \item \textbf{Dataset Limitations:} The anemia screening model was trained on a relatively small dataset (n=250), limiting generalizability across diverse populations, especially with varying skin tones and lighting conditions.
    
    \item \textbf{Diagnostic Scope:} The current implementation focuses solely on anemia screening. Broader diagnostic capabilities would require additional data and validation.
    
    \item \textbf{Hardware Constraints:} Although the system performs well on Jetson Nano, it is still a device that does not represent the lower-cost end of embedded devices.
    
    \item \textbf{Field Validation:} The system has yet to undergo clinical validation in real-world field deployments, which is critical to assess usability, reliability, and impact in operational healthcare settings.
\end{enumerate}

Our foundational work expands on the development of a broader class of constraint-resilient, edge-powered health support systems. We aim to support the global mission of improving healthcare accessibility and reducing health disparities worldwide. Therefore, we plan to extend our work in the following key areas:

\begin{itemize}
    \item \textbf{Multimodal Physiological Sensing:} Expanding the platform to support additional screening modules (e.g., blood pressure, oxygen saturation, malnutrition screening) while maintaining system efficiency.

    \item \textbf{Hardware Diversification:} Exploring deployment on emerging low-cost edge platforms (e.g., Coral Edge TPU, ESP32 AI modules) to further reduce hardware costs.

    \item \textbf{Energy Harvesting \& Sustainability:} Investigating the integration of renewable energy sources (e.g., solar-powered kits) to enhance system autonomy in off-grid environments.
\end{itemize}

\section{Acknowledgements}
This work is supported by the Center for Research \& Development at the University of Puerto Rico at Mayagüez and
the NSF-EPSCoR Center for the Advancement of Wearable Technologies, NSF grant OIA-1849243.


\end{document}